\documentclass[fleqn,usenatbib]{mnras}

\usepackage[T1]{fontenc}
\usepackage{ae,aecompl}

\usepackage{xspace}
\usepackage{graphicx}
\usepackage{amsmath}
\usepackage{amssymb}

\newcommand{\MJup}{\ensuremath{M_{\mathrm{Jup}}}\xspace}

\newcommand{\Teff}{\ensuremath{T_{\rm eff}}\xspace}
\newcommand{\logg}{\ensuremath{\log\,g}\xspace}

\newcommand{\mic}{$\mu$m\xspace}
\newcommand{\as}{\hbox{$^{\prime\prime}$}\xspace}
\newcommand{\lsd}{\hbox{$\lambda/D$}\xspace}

\title[High-contrast imaging of Sirius~A with SPHERE]{High-contrast imaging of Sirius~A with VLT/SPHERE: \\ Looking for giant planets down to one astronomical unit\thanks{Based on observations made with ESO Telescopes at the La Silla Paranal Observatory under program IDs 60.A-9365 and 60.A-9382.}}

\author[A. Vigan et al.]{
A. Vigan,$^{1,2}$\thanks{E-mail: \href{mailto:arthur.vigan@lam.fr}{arthur.vigan@lam.fr}}
C. Gry,$^{1}$
G. Salter,$^{1}$
D. Mesa,$^{3}$
D. Homeier,$^{4,5}$
C. Moutou$^{1,6}$ and
\newauthor
F. Allard$^{5}$ \\
$^{1}$Aix Marseille Universit\'e, CNRS, LAM (Laboratoire d'Astrophysique de Marseille) UMR 7326, 13388, Marseille, France \\
$^{2}$European Southern Observatory, Alonso de C\'ordova 3107, Vitacura, Cassilla 19001, Santiago, Chile \\
$^{3}$INAF - Osservatorio Astronomico di Padova, Vicolo dell'Osservatorio 5, 35122 Padova, Italy \\
$^{4}$Zentrum f{\"u}r Astronomie der Universit{\"a}t Heidelberg, Landessternwarte K\"onigstuhl 12, 69117 Heidelberg, Germany \\
$^{5}$\'Ecole Normale Sup\'erieure, Lyon, CRAL (UMR CNRS 5574), Universit\'e de Lyon, 69364 Lyon Cedex 07, France \\
$^{6}$CNRS, Canada–France–Hawaii Telescope Corporation, 65-1238 Mamalahoa Hwy., Kamuela, HI 96743, USA
}

\date{Accepted 2015 August 18.  Received 2015 August 6; in original form 2015 June 20}

\pubyear{2015}

\begin{document}
\label{firstpage}
\pagerange{\pageref{firstpage}--\pageref{lastpage}}
\maketitle

\begin{abstract}
Sirius has always attracted a lot of scientific interest, especially after the discovery of a companion white dwarf at the end of the 19$^{\rm th}$ century. Very early on, the existence of a potential third body was put forward to explain some of the observed properties of the system. We present new coronagraphic observations obtained with VLT/SPHERE that explore, for the very first time, the innermost regions of the system down to 0.2\as (0.5~AU) from Sirius~A. Our observations cover the near-infrared from 0.95 to 2.3~\mic and they offer the best on-sky contrast ever reached at these angular separations. After detailing the steps of our SPHERE/IRDIFS data analysis, we present a robust method to derive detection limits for multi-spectral data from high-contrast imagers and spectrographs. In terms of raw performance, we report contrasts of 14.3~mag at 0.2\as, $\sim$16.3~mag in the 0.4--1.0\as range and down to 19~mag at 3.7\as. In physical units, our observations are sensitive to giant planets down to 11~\MJup at 0.5~AU, 6--7~\MJup in the 1--2~AU range and $\sim$4~\MJup at 10~AU. Despite the exceptional sensitivity of our observations, we do not report the detection of additional companions around Sirius~A. Using a Monte Carlo orbital analysis, we show that we can reject, with about 50\% probability, the existence of an 8~\MJup planet orbiting at 1~AU.
\end{abstract}


\section{Introduction}
\label{sec:introduction}

As the brightest star in the sky, Sirius has attracted the attention of humanity since prehistoric times. After the discovery of a second component in the Sirius system in 1862 \citep{bond1862}, 18 years after its theoretical prediction \citep{bessel1844}, the system has been extensively monitored using various observing techniques.

In addition to the discovery of Sirius B as a white dwarf companion, there are particularities of the Sirius system that continue to generate much scientific interest. Sirius~A and B form a close binary system, with a 50 year period \citep{vandenbos1960}, on an elliptical orbit ($e=0.59$) that brings Sirius~B closer to Sirius~A (8~au) than Saturn from the Sun. Sirius~A is known to have the earliest spectral type amongst white dwarf companions \citep{holberg2013}, and Sirius~B has one of the highest known masses for a white dwarf \citep{gatewood1978}.

Ancient Greek and Chinese records also suggest that the colour of Sirius may have changed within historical times \citep{gry1990}. One of the possible scenarios put forward to explain this change in colour is related to a transient reddening of Sirius caused by matter ejected by tidal interaction with a potential third body \citep{bonnet-bidaud1991}. Irregularities in the orbital motion of Sirius~B were detected very early on suggesting the existence of a third component in the system. Detection of possible visual companions \citep[e.g.][]{baize1931} and of a 6.3 years periodicity perturbation in the orbit of the Sirius~A--B system \citep{benest1995} has fed hopes of detecting a low mass companion in the system. In particular, the periodic perturbation reported by \citet{benest1995} led them to predict the existence of a companion of maximum mass $\sim$50~\MJup orbiting Sirius~A at less than 3\as (7.9~au). Because of the relatively young age of Sirius~A \citep[225--250~Myr;][]{liebert2005}, a companion of this mass would be easily detectable with imaging in the near-infrared \citep[e.g.][]{burrows1997,baraffe2003}. 

These particularities make the search for stellar or sub-stellar companions to Sirius~A, or B, with imaging particularly attractive. However, the extreme brightness of Sirius~A has long been known to represent a major issue for such a search. Several attempts at detecting possible companions by imaging the close stellar field around Sirius from the ground using coronagraphic devices \citep{bonnet-bidaud1991,bonnet-bidaud2000} and adaptive optics \citep[ESO/ADONIS;][]{bonnet-bidaud2008}, or from space with the HST \citep{schroeder2000}, have proven unsuccessful. Deep imaging performed with ADONIS in $K$~band reached a limiting contrast of 9.5~mag at 3\as to 13.1~mag at 10\as from Sirius~A,  this translates to the ability to detect massive brown dwarf companions. The most recent and most sensitive study to date was performed by \citet{thalmann2011} using Subaru/IRCS at 4.05~\mic in saturated imaging, which allowed them to achieve a sensitivity to companions of 6--12~\MJup at 1\as, 2--4~\MJup at 2\as and 1.6~\MJup beyond 4\as. While they provide very strong constraints against the existence of the massive companion with a 6.3--yr period predicted by \citet{benest1995}, their sensitivity to giant planets at very small angular separation ($\le$0.7\as, 1.8~au) is limited by the saturation of Sirius~A on the detector. Thalmann et al. (2011) conclude their analysis with the hope that new-generation high-contrast imagers would provide tighter constraints at very small separations.

Four years later, we report the first observations of Sirius~A with SPHERE \citep[Spectro-Polarimetric High-contrast Exoplanet REsearch;][]{beuzit2008}, the new generation high-contrast imager for the Very Large Telescope (VLT). We use this new instrument to perform the closest and deepest imaging of (sub-)stellar companions around Sirius~A. Apart from the strong scientific interest of potentially detecting either a stellar or planetary-mass object around Sirius~A, these observations play the important role of demonstrating SPHERE's capability to observe nearby regions to the brightest stars.

In Section~\ref{sec:observations} we present in more details the instrument and the observations. Section~\ref{sec:data_reduction} details the data reduction steps performed to accurately calibrate the data. In particular, we present several useful procedures for the new community of SPHERE users. In Section~\ref{sec:speckle_noise_attenuation} we perform the high-contrast analysis of our data to try to reveal the presence of faint companions, and in Section~\ref{sec:determination_detection_limits} we detail a new procedure to derive reliable detection limits for SPHERE near-infrared data, comparing it to more a straightforward approach. Finally, we discuss our results and their implications in Section~\ref{sec:discussion}, before detailing our conclusions in Section~\ref{sec:conclusions}.

\section{Observations}
\label{sec:observations}

\begin{table*}
  \caption{Observing log on 2014 December 07}
  \label{tab:observing_log}
  \centering
  \begin{tabular*}{1.0\linewidth}{cccc@{\extracolsep{\fill}}ccccc}
  \hline\hline  
       &                & \multicolumn{2}{c}{IFS}                        & \multicolumn{2}{c}{IRDIS}                                &          &                         &        \\\cline{3-4}\cline{5-6}
  Seq. & LST time       & DIT$\times$NDIT$\times$NEXP & T$_{\mathrm{exp}}$ & DIT$\times$NDIT$\times$N$_{\rm dither}$ & T$_{\mathrm{exp}}$ & FoV rot. & Seeing$^{\rm a}$ & Sr$^{\rm a}$ \\
       & (hr:min)       &                             & (min)            &                                       & (min)            & (deg)    & (arcsec)                    & (\%)   \\
  \hline                                                                  
  1 & 04:41--05:13  & 2$\times$34$\times$16       & 18               & 4$\times$25$\times$16                 & 27               & 4        & $0.52 \pm 0.10$ & $92.8 \pm 1.6$ \\
  2 & 05:23--05:32  & 2$\times$34$\times$4        &  5               & 4$\times$25$\times$4                  &  7               & 2        & $0.46 \pm 0.05$ & $93.9 \pm 0.5$ \\
  3 & 05:55--06:29  & 2$\times$34$\times$16       & 18               & 4$\times$25$\times$16                 & 27               & 29       & $0.35 \pm 0.05$ & $95.0 \pm 0.6$ \\
  4 & 06:37--07:11  & 2$\times$34$\times$16       & 18               & 4$\times$25$\times$16                 & 27               & 50       & $0.42 \pm 0.09$ & $94.5 \pm 1.0$ \\
  \hline
  all &                 &                             & 59               &                                       & 119              & 106      & $0.43 \pm 0.10$ & $94.1 \pm 1.4$ \\
  \hline
  \end{tabular*}
\begin{flushleft}
$^{\rm a}$ The seeing and Strehl ratio estimations are calculated over periods of 10~s every 30~s by the real-time computer. The Strehl ratio is expressed at 1.6~\mic. The error bar is calculated as the standard deviation of the values.
\end{flushleft}
\end{table*}

Sirius~A has been observed as part of the SPHERE instrument science verification under ESO programme ID~60.A-9365 (P.I. Gry). The SPHERE, instrument, installed at the VLT \citep{beuzit2008}, is highly specialized being dedicated to the high-contrast imaging and spectroscopy of young giant exoplanets. It is based on the SAXO extreme adaptive optics (AO) system \citep{fusco2006,petit2014,sauvage2014}, which controls a $41\times41$ actuator deformable mirror, and 4 control loops (fast visible tip-tilt, high-orders, near-infrared differential tip-tilt and pupil stabilization). The common path optics include several stress polished toric mirrors \citep{hugot2012} to transport the beam to the coronagraphs and scientific instruments. Several types of coronagraphic devices for stellar diffraction suppression are provided, including apodized pupil Lyot coronagraphs \citep{soummer2005} and achromatic four-quadrants phase masks \citep{boccaletti2008}. The instrument is equipped with three scientific subsystems: the differential imager and spectrograph \citep[IRDIS;][]{dohlen2008a}, an integral field spectrograph \citep[IFS;][]{claudi2008} and the ZIMPOL rapid-switching, imaging polarimeter which works in the visible \citep{thalmann2008}.

Observations presented here were acquired on the 2014 December 07 in the \texttt{IRDIFS\_EXT} mode, where both the IFS and IRDIS observe in parallel. In this setup, the IFS covers wavelengths from 0.95 to 1.65~\mic (Y--H~bands) at a resolution of $\sim$30 in a $1.7\arcsec\times1.7\arcsec$ square field of view (FoV), while IRDIS observes in the dual-band imaging mode \citep[DBI;][]{vigan2010} with K12, a pair of filters in the $K$~band ($\lambda_{\rm K1} = 2.110$~\mic, $\lambda_{\rm K2} = 2.251$~\mic, $\sim$0.1~\mic bandwidth), within a $\sim$$4.0\arcsec$ radius circular FoV. The observations were performed in pupil-stabilized mode to perform angular differential imaging \citep[ADI;][]{marois2006} with an apodized pupil Lyot coronagraph designed for the $K$~band (\texttt{ALC\_Ks}), which uses a coronagraphic mask of radius 120~mas and an optimized Lyot stop. The IRDIS detector was dithered on a 4$\times$4 pattern to reduce the effect of the residual flat-field noise.

Originally the programme was allocated two hours to observe Sirius~A and B, each for one hour. Unfortunately, the observations of Sirius~B could not be performed due to the brightness of Sirius~A, located 9.9\arcsec away at the time of the observations. During the observations, bright stray light diffracted by the telescope spider vanes entered the FoV of the wavefront sensor, causing an over-illumination and an AO loop crash, and prevented further observations of Sirius~B. To take advantage of the full time allocation of the programme, the sequence on Sirius~A was then extended after the failed attempt on Sirius~B.

The complete sequence on Sirius~A spanned 2 h and 30 min, totalling 106~deg of FoV rotation. The time was divided into four continuous sequences, with interruptions of 10, 23 and 8~min in-between, the longest interruption corresponding to the attempt at observing Sirius~B. For IRDIS, the total integration time was 88~min, divided into data cubes of DIT$\times$NDIT\footnote{DIT: detector integration time; NDIT: number of DIT.}=4$\times$25~s, while for the IFS it was 59~min, divided into data cubes of DIT$\times$NDIT=2$\times$34~s. The difference in total integration time between the two instruments is only the result of detector readout overheads that are more important for the IFS (efficiency of 51\% for DIT=2~s) than for IRDIS (efficiency of 77\% for DIT=4~s). The conditions were excellent, with seeing values below 0.5\arcsec for the duration of the observations. The observing conditions as reported by the observatory were clear. Combined with the extreme brightness of Sirius~A ($R = -1.46$), this resulted in exquisite AO performance with reported Strehl ratios of over 90\% for all observations. For sequences 1, 2 and 4, the AO spatial filter was set to a size of 1.31~\lsd, and for sequence 3, during the best observing conditions, the size of the spatial filter was decreased to an even smaller size of 1.14~\lsd. The coronagraphic observations of Sirius~A are summarized in Table~\ref{tab:observing_log}.

The time between long coronagraphic observations was used to acquire calibration data that will be used for the data analysis. These data include the following:

\begin{itemize}
  \item Stellar point-spread functions (PSFs) taken off-axis ($\sim$0.4\as) with the neutral density (ND) filter ND3.5, which provides the highest attenuation factors in SPHERE -- approximately 3\,300 in $Y$~band, 16\,300 in $J$~band, 1\,200 in $H$~band, and 700 in $K$~band. For this calibration, the star is moved away from the coronagraph by applying an offset on the near-infrared differential tip-tilt plate. During this observation, the AO visible tip-tilt and high-order loops remain closed to provide a diffraction-limited PSF. Three off-axis PSFs were acquired, one at the beginning of sequence 1 and two at the beginning of sequence 3.
  \item `Star centre' coronagraphic images where four symmetric satellite spots are created by introducing a periodic modulation on the deformable mirror. This data is used in subsequent analysis to determine an accurate position of the star centre behind the coronagraph, and hence the centre of field rotation. Four star centre frames were acquired, one during each of the observing sequences.
  \item Sky backgrounds taken at a distance of $\sim$30\as from Sirius in a region of the sky empty of stars. These backgrounds are necessary for the IRDIS observations in $K$~band, where the sky has a significant contribution.
\end{itemize} 

Data from the SPARTA real-time computer of the SAXO extreme AO system were collected at regular intervals in parallel of all the observations, including images from the differential tip-tilt sensor (DTTS). The DTTS removes a minute fraction of the incoming flux in the near-infrared (at 1.5~\mic) to image the PSF just before the coronagraph. This it used for the DTTS loop which keeps the PSF locked on the coronagraph once the observing sequence has started. These files also contain estimations of the seeing, wind speed and Strehl ratio made by the real-time computer.

Standard calibrations for the \texttt{IRDIFS\_EXT} mode were acquired in the morning as part of the instrument calibration plan. For the IFS, this includes deep instrumental backgrounds with the same DIT values as the science data, detector flats at four different wavelengths and with a white lamp, an integral field unit (IFU) flat that allows to calibrate the response of each individual lenslet, and finally a wavelength calibration frame where the IFU is illuminated using four different laser lines. For IRDIS, this includes deep instrumental backgrounds and a detector flat-field acquired in the K12 filter pair.

\section{Data reduction}
\label{sec:data_reduction}

In this section we describe the data reduction of our Sirius observations, i.e. the steps that we followed to go from the raw data to cleaned and aligned data cubes that will later be used for our scientific analysis. Due to several issues with this data set, we choose to detail the data reduction quite extensively. We provide useful information that could be used for the reduction of SPHERE data by other users. In Appendix~\ref{sec:improving_irdifs_data_reduction}, we describe some of the more technical details and we provide a link to our SPHERE/IFS pre-processing pipeline.

We call pre-processing the steps of creating the basic calibrations and applying them to the science data. We distinguish the case of IFS and IRDIS data because they are two conceptually very different instruments. For both instruments, we made use of the preliminary release (v0.14.0-2) of the SPHERE data reduction and handling (DRH) software \citep{pavlov2008}, as well as additional tools created for the purpose of improving or replacing the DRH.

\subsection{IFS pre-processing}
\label{sec:ifs_preprocessing}

For the IFS, we followed the steps described in \citet{mesa2015} for the creation of some of the basic calibrations using the DRH: backgrounds, master flat-field, IFS spectra positions, initial wavelength calibration and IFU flat-field. After these calibrations, one would normally use directly the DRH to produce the $(x,y,\lambda)$ data cubes, but we introduce an additional pre-processing that performs the following actions:

\begin{itemize}
  \item accurate determination of the time and parallactic angle of each science frame based on header information (\texttt{DATE-OBS} and \texttt{DATE} FITS keywords for the start and end of each data cube);
  \item subtraction of the background to each science frame;
  \item normalization of the data by the value of the DIT and by the attenuation of the ND filter used using their transmission curves\footnote{\url{https://www.eso.org/sci/facilities/paranal/instruments/sphere/inst/filters.html}};
  \item temporal binning of the raw science data to reduce the total number of frames. Sirius~A, being extremely bright, required the minimum DIT of 2~s to be used, resulting in more than 1750 independent science frames. The binning was performed by directly averaging groups of four consecutive frames. Given the excellent observing conditions and the overall quality of the data, we did not have to remove any bad frames, such as AO open-loops. The binning was performed within each of the data cubes, i.e. we did not bin frames coming from two different cubes. This resulted in the loss of $\sim$150 frames over the whole sequence because the NDIT of the cubes was not a multiple of 4. After the binning, the number of frames was reduced from 1750 frames to 400. The maximum field rotation per binned frame is equal to 0.38$^{\rm o}$ at transit. It is important to perform the binning of the science data at the level of the raw science data rather than at a later stage after multiple interpolations, so as to avoid the addition of interpolation noise. During the binning process, the timing and position angle of each binned frame is determined using the information of each individual raw frame;
  \item correction of bad pixels identified using the master dark and master flat-field DRH products -- this step is introduced as a substitute to the bad pixel correction provided by the DRH. This cosmetic step is particularly important due to the large number of isolated bad pixels in the IFS detector. We compared several methods and found that the \texttt{MASKINTERP} IDL procedure\footnote{\url{http://physics.ucf.edu/~jh/ast/software.html}}, performing a bicubic pixel interpolation using neighbouring good pixels as reference, gave the best results;
  \item cross-talk correction. The light passing through the IFS lenslets forms one PSF for each lenslet. However, part of the light at each PSF position is actually contaminated by spurious signal from adjacent lenslets. Moreover, a second effect for the IFS spectra is that spurious light coming from other wavelengths can also contaminate the light at a considered wavelength. We refer the reader to the work of \citet{antichi2009} for a theoretical treatment of the SPHERE IFS crosstalk. The consequence of this cross-talk effect on the IFS calibrated data cubes is the presence of spurious light from different wavelengths in each spectral frame. This effect is particularly evident when looking at images with the satellite spots used to centre each frame. In this case, a double satellite spot can be seen especially when looking to wavelengths where the total flux from the star is very low. We have developed a two-step procedure, that is applied to the raw data, to solve this problem. The first step corrects the small scale cross-talk using a kernel of $41\times41$~pixels that contains a Moffat function of the type $1/(1+r_{\mathrm{dist}}^3/b_{\mathrm{fac}}^3)$, where $r_{\mathrm{dist}}$ is the distance from the centre of the array (expressed in pixels) and $b_{\mathrm{fac}}$ is an appropriate numerical value that has been determined iteratively by trying to minimize the brightness of the secondary satellite spots present at particular wavelengths as described above. The data is first convolved with the kernel, and the convolved image is then subtracted from the original data. To avoid subtracting the value of the central pixels, the $3\times3$~pixel central region of the array is set to zero. The second step consists of removing the large-scale cross-talk that corresponds to the instrumental background in $64\times64$~pixel boxes. This step is not necessary when subtracting a proper instrumental background as we do in the second step of the intermediate pre-processing.
\end{itemize}

The partially calibrated, temporally binned, frames are then injected into the DRH recipe that creates the data cubes by interpolating the data spectrally and spatially. In this step, no background subtraction or bad pixel correction are performed, since these calibrations have already been made.

As mentioned previously, the IFS detector is strongly affected by bad pixels. Even with a thorough initial cleaning of the raw science frames, the produced $(x,y,\lambda)$ data cubes still show remaining bad pixels, mostly in the low signal-to-noise ratio (SNR) wavelengths corresponding to atmospheric absorption bands. To finish the pre-processing of the science data, these bad pixels are first identified using a sigma-clipping routine, and then corrected using the \texttt{MASKINTERP} procedure. 

An important additional step that we have implemented is a correction of the wavelength calibration, as performed by the DRH, which reduces calibration errors on from $\sim$20~nm (at worst) to 1.4~nm. As this procedure is useful for the exploitation of the instrument by the community, we describe it in detail in Appendix~\ref{sec:correction_ifs_wavelength_calibration}. As a byproduct of the wavelength calibration step we also obtain the spatial rescaling factor that will later be used to analyse the data (see Section~\ref{sec:speckle_noise_attenuation}).

\subsection{IRDIS pre-processing}
\label{sec:ird_preprocessing}

Pre-processing of IRDIS data is much more straightforward owing to its simple optical concept. This part of the analysis is handled by the LAM-ADI pipeline (\citealt{vigan2012}, Vigan et al. submitted), which has been updated to handle multi-wavelength data from IRDIS and the IFS. Each of the images in the coronagraphic observing sequences are background subtracted and divided by the flat-field. Bad pixels are corrected using the bad pixel maps created by the DRH which replaces them with the median of neighbouring good pixels. Since the IRDIS DBI mode uses well-characterized filters, the scaling factor between the two wavelengths is taken as the ratio of the central wavelength of the two filters. We also apply some temporal binning of the data by averaging groups of four consecutive frames, resulting in a reduction of the number of frames from 1300 to 306. The maximum field rotation per binned frame at transit is 0.52$^{\rm o}$.

\subsection{Off-axis reference PSF}
\label{sec:off_axis_reference_psf}

Sirius~A is the brightest star in the night sky, making its observation difficult without a coronagraph. The ND3.5 filter used for our observations (see Section~\ref{sec:observations}) was not sufficient to completely avoid saturation of the off-axis PSF of Sirius~A. The PSF was slightly saturated or out of the linear range in the first five and last four spectral channels of the IFS, and the core of the PSF in the K12 filters was saturated in IRDIS over a diameter of $\sim$\lsd.

To solve this problem, we make use of an observation of $\beta$~Pictoris performed the day after, also part of the SPHERE science verification under ESO programme ID~60.A-9382 (P.I. Lagrange). This target was observed in the same instrumental configuration (\texttt{IRDIFS\_EXT}) and in extremely similar conditions (seeing of $0.30\arcsec \pm 0.03\as$, Strehl ratio of $94.6\% \pm 0.5\%$, clear observing conditions). The off-axis PSF was acquired using the same ND3.5, but without any saturation (maximum number of counts at $\sim$12\,000, well below the linearity limit of 20\,000). The data for the $\beta$~Pictoris off-axis PSFs from the IFS and IRDIS is calibrated and pre-processed in exactly the same manner as described in Sections~\ref{sec:ifs_preprocessing} and \ref{sec:ird_preprocessing}. Then, this proxy PSF is scaled in flux to properly represent the Sirius PSF. Our scaling procedure is described in Appendix~\ref{sec:scaling_off-axis_proxy_psf}. We estimate an uncertainty on the final scaled PSF of 0.06~mag for IRDIS, and $\sim$0.2~mag for the IFS modified spectral channels.

\subsection{Centring of the coronagraphic data}
\label{sec:centring_coronagraphic_data}

During the observations of Sirius~A, two of the atmospheric dispersion correctors (ADCs) were not tracking, one in the visible and one in the near-infrared. There are two major consequences to this malfunction: (1) there is residual dispersion at the level of the coronagraph, which means that the star is not well centred behind the mask at all wavelengths, and (2) this centring is evolving with time as the other ADC of each pair continues tracking while its counterpart does not move. The first effect is mitigated by the fact that SPHERE uses a DTTS in its near-infrared arm to compensate the differential refraction between the visible (where the wavefront sensing is performed) and the near-infrared. Due to the DTTS, whatever happens with the ADCs, the star will remain centred on the coronagraph at the wavelength of the DTTS. This is very important since it provides us with a reference wavelength for the centring.

For the IFS, we verify that this is indeed the case by measuring the position of the centre at every wavelength for each of the three star centre frames that were acquired\footnote{One was acquired at the end of observing sequence 1, one 15 min later at the beginning of sequence 2, and one 1\,h\,15\,min later at the beginning of sequence 3.} on Sirius~A. As expected, at around 1.5~\mic the centres are within 0.5~pixel (3.5~mas) of each other, while at 1.0~\mic they are separated by more than 2~pixels. In addition, a visual assessment of the coronagraphic centring shows it to be accurate at longer wavelengths. This is expected as the original centring is performed in the $H$~band. We adopt as the reference wavelength the spectral channel number 35 ($\lambda_{35} = 1.58$~\mic), and we recentre all the other channels relative to it, as described in Appendix~\ref{sec:centring_ifs_coronagraphic_data}.

The analysis is much simpler for IRDIS, due to the differential refraction attenuating with increasing wavelength. The three centres determined using the star centre frames are all within 0.16~pixel (2~mas), which means that the star centring on the coronagraph remained very stable throughout the full observation, despite the ADC tracking problem. A visual assessment of the data shows that the absolute centring of the star behind the coronagraph appears satisfactory, with no apparent decentring. For the IRDIS sequence, we adopt as the centre for the whole sequence the average of the three centres calibrated from the star centre frames. No additional recentring was required for IRDIS.

\section{Speckle noise attenuation}
\label{sec:speckle_noise_attenuation}

\begin{figure*}
  \centering
  \includegraphics[width=\linewidth]{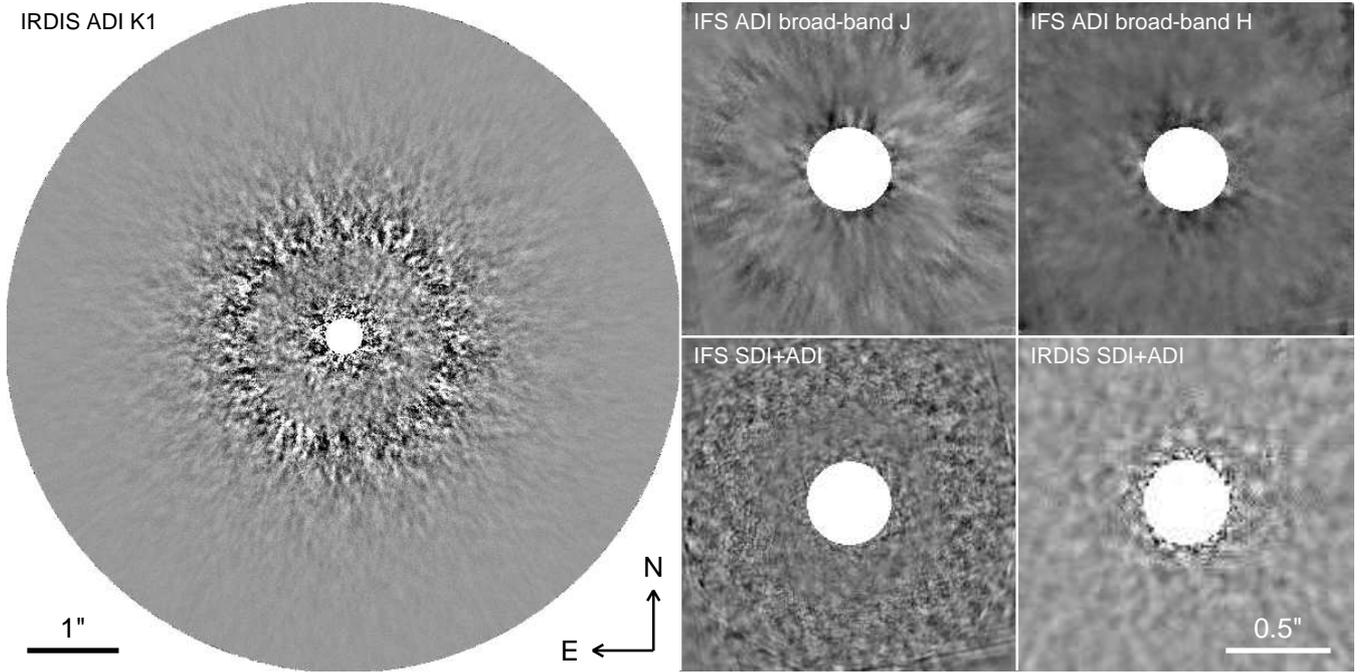}
  \caption{Final images obtained after ADI or SDI+ADI analysis of the IFS and IRDIS data. Left: IRDIS FoV after ADI analysis in the K1 filter. Right, top row: IFS FoV after ADI analysis of the data spectrally binned to equivalent $J$ and $H$ broad-band filters. Right, bottom row: IFS and IRDIS inner FoV after the full SDI+ADI analysis. The speckles and stellar halo have been subtracted using a PCA analysis (see text for details about the analysis).}
  \label{fig:irdifs_images}
\end{figure*}

After performing the calibrations described in previous sections, we are able to process our data using classical high-contrast imaging analysis techniques. The goal of these techniques is to attenuate or remove the stellar halo and speckle pattern that limit the ability to detect faint companions. For this purpose, we use both the angular diversity provided by ADI \citep{marois2006} and the spectral diversity provided by spectral differential imaging \citep[SDI;][]{racine1999}. Over the years, many data analysis algorithms have been developed to take advantage of such diversity. We use as our primary means of investigation a principal component analysis (PCA) following the formalism described by \citet{soummer2012}. Because we have multi-dimensional data (temporal and spectral), there are several possibilities that can be used to build the PCA modes and subtract the speckle pattern. Below we describe the three ways we use to analyse our Sirius data set.

\subsection{Angular differential imaging analysis}

The first technique consists of looking at each spectral channel of the IFS and IRDIS independently and only apply ADI to remove the speckle pattern. For any given spectral channel, the PCA modes are constructed using all images of this channel, resulting in a total of 400 modes for the IFS and 306 for IRDIS. The modes are calculated over the whole image, from an inner radius of 0.2\as, up to 0.8\as for the IFS and up to 3.7\as for IRDIS. While not necessarily the absolute best approach in terms of performance, the quality and stability of the data ensures excellent results with this scheme, while keeping the computing time reasonable. An increasing number of modes (up to 30\% of the total number) is subtracted to each of the science frames, before applying derotation and averaging all the speckle-subtracted images.

A variant of this method, that we also explore, is to bin the data spectrally before applying ADI. For the IFS we bin the data into the equivalent $Y$, $J$ and $H$ broad-band filters, and apply the same analysis as described above. For IRDIS, we simply add the K1 and K2 data to form a single image. 

These methods are equivalent to considering that we have independent data sets at different wavelengths. While it is efficient for the subtraction of the speckles, it does not take advantage of the spectral dimension to further reduce the level of speckle-noise. 

\subsection{Angular and spectral differential imaging}
\label{sec:adi+sdi}

Our third method uses the spectral information to perform SDI in addition to ADI. The PCA modes are constructed using all the frames, after rescaling them spatially (see Appendices~\ref{sec:correction_ifs_wavelength_calibration} and \ref{sec:spatial_scaling_algorithm}) in order to make the speckle pattern match at all wavelengths. In the rescaled images, the speckle pattern is stable while the signal of a companion is moving with time and wavelength. This is more aggressive in terms of speckle subtraction, but the large spectral range covered by the IFS ($Y$--$H$) gives room for using SDI with a safe margin. The bifurcation radius \citep{thatte2007} is equal to 148~mas. This defines the separation above which the signal of a companion will have moved by more than \lsd after the spatial rescaling. Above this radius we will be able to build a reference PSF that is not significantly biased by that of the companion. With this approach, we obtain 15\,600 modes for the IFS data set, and 612 modes for IRDIS. Here again, the modes are calculated over the whole image. For the IFS analysis, we subtract up to 5\,000 modes in steps of 50 modes, and for IRDIS we subtract up to 500 modes in steps of 10 modes. After the modes subtraction, all frames are spatially rescaled back to their original size, derotated and averaged to obtain the final image.

\subsection{Results}

The three analyses are performed from an inner radius of 0.2\as for both the IFS and IRDIS. This inner radius is not extended down to the theoretical inner-working angle of the coronagraph, 120~mas, due to the ADC tracking problem. After the recentring of the frames with respect to the real star centre (see Section~\ref{sec:centring_coronagraphic_data}), the coronagraphic mask is offset from the centre of the images. We estimate that, in the worst case, the coronagraphic mask covers up to 0.18\as. We add a small margin and provide quantitative measurements from 0.2\as outwards.

All the final images were inspected visually and compared to each other to look for point-like structures. The multi-wavelength coverage was used to discriminate any speckle residual from the signal of any potential real detection. We have not identified any point-like structure in any of the images and all visible structures in the images have been identified as residual speckle noise by our multi-wavelength comparison. Fig.~\ref{fig:irdifs_images} presents several of the final images obtained with either the IFS or IRDIS. Although some structures are still visible after subtraction of the speckles, none of them can be identified as a real detection.

\section{Determination of the detection limits}
\label{sec:determination_detection_limits}

\subsection{Method description}
\label{sec:detlim_method_description}

Deriving detection limits in high-contrast imaging is getting increasingly difficult as the new generation of instruments incorporate high-performance dual-band imagers, such as IRDIS, or IFSs. The determination of detection limits in DBI data in the case of no-detection has already been studied previously in the context of the SPHERE/IRDIS scientific preparation or surveys with VLT/NACO in SDI mode (\citealt{vigan2010}; \citealt{maire2014}; Rameau et al. submitted). There are essentially two approaches that have been explored so far: (1) the semi-analytical approach where the residual flux after SDI reduction is predicted as a function of separation and physical parameters of a planet using the evolutionary model tables, allowing a direct conversion of the residual noise in the speckle-subtracted images into mass/temperature (Rameau et al. submitted); (2) the data analysis approach where one introduces in each image fake planets with flux based on simulated spectra \citep{maire2014}. Whatever the approach, it is generally agreed that the result of data analysis involving SDI cannot be directly converted to a physical mass limit because of the degeneracy induced by the spectral difference.

For IFS data, the relative novelty of this type of instrumentation in high-contrast imaging has not yet triggered extensive studies about how to best derive accurate physical detection limits. This is however a very important problem as all large direct imaging surveys that recently started using IFSs will require the derivation of such limits for a large number of stars to be able to perform the final statistical analysis of the giant planet population. 

We present here a method that we have developed for VLT/SPHERE IRDIFS data, based on the second approach described above for SDI data, but modified for our simultaneous IRDIS and IFS data. In summary, we inject fake planets into the data, perform the speckle removal analysis, and then measure the level at which each of the planets is detected to establish a detection limit. To make the analysis more efficient, several fake planets are injected simultaneously in each analysis at the same contrast level. They are placed at increasing position angles and separation from the star to form a spiral. Increments in position angle and separation are chosen so as to make sure that the self-subtraction effects, created radially by SDI and azimuthally by ADI, do not overlap for different fake planets. The fake planet pattern is identical between the IFS and IRDIS, and for each contrast level the pattern is introduced at 10 different orientations to improve the final detection statistics. The smearing induced by the FoV rotation in each of the frames is properly taken into account when creating the fake planets map.

For the speckles subtraction, we perform for each newly-created data set the same analysis as described in Section~\ref{sec:speckle_noise_attenuation} using the SDI+ADI approach. At the end of each analysis, the final image is convolved with a \lsd aperture, in order to remove noise at spatial scales smaller than the expected planet signal. Then we define the SNR of each of the injected planet as the maximum value of the convolved image at the known location of the planet divided by the rms of statistically independent pixels (i.e. separated by more than \lsd) in an annulus located at the same separation as the planet in the convolved image. Within this annulus, we exclude the signal of the planet itself, the area located at $\pm25\deg$ on either side that contains significant self-subtraction effects, and the radial spectral self-subtraction effects from other fake planets located at other separations and position angles. 

We perform two distinct analyses that serve different purposes. In the first, detailed in Section~\ref{sec:flat_spectrum}, we introduce fake planets with a stellar-like spectrum, i.e. with the same contrast ratio with respect to the star in each spectral channel. In the second, detailed in Section~\ref{sec:simulated_spectra}, the contrast of the fake planets is based on synthetic spectra representing planets at different masses.

Our analysis focuses only on the \emph{detection} of companions, as opposed to their \emph{characterization}. Indeed, these are two distinct aspects of the same problem that do not require the same treatment. While the data analysis focused on characterization depends on the accurate calibration of all the biases on the photometry and astrometry, data analysis focused on detection can afford to be much more aggressive, at the expense of quantitative flux measurements.

\subsection{Raw contrast detection limits using fake planets with stellar-like spectra}
\label{sec:flat_spectrum}

\begin{figure}
  \centering
  \includegraphics[width=\columnwidth]{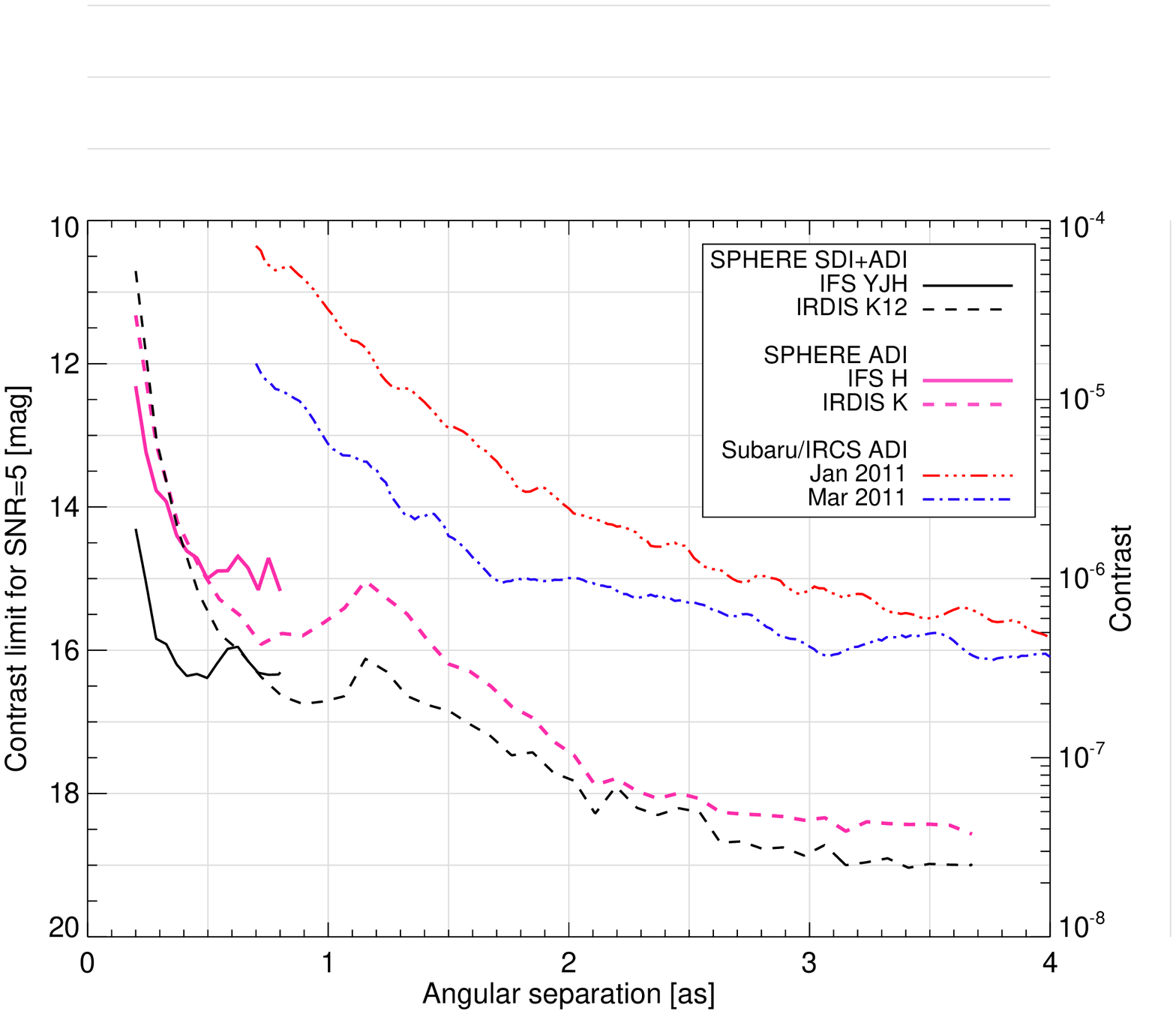}
  \caption{SPHERE sensitivity in contrast to obtain a detection at SNR=5 with the IFS (plain curve) and IRDIS (dashed curve) assuming a companion with a constant contrast ratio with respect to the star in all spectral channels. We show the sensitivity obtained both for an SDI+ADI analysis and for an ADI-only analysis (see text for details). The self-subtraction effects at small separation in IRDIS DBI mode strongly affect the performance below 0.7\arcsec, while the wide spectral coverage of the IFS (0.96--1.6~\mic) allows to mitigate this effect and explore smaller separations. These limits are compared to the limits published by \citet{thalmann2011} with Subaru/IRCS in the Br~$\alpha$ filter at 4.05~\mic.}
  \label{fig:detection_limit_mag}
\end{figure}

Introducing fake planets with a stellar-like spectrum, i.e. with a constant contrast ratio with respect to the star, is not physically realistic, but it is useful to get an idea of the raw performance, which can be compared between instruments, since it is in no way model-dependent. It is, however, pessimistic when using SDI because the self-subtraction effects at small angular separation will significantly affect the sensitivity. 

Fig.~\ref{fig:detection_limit_mag} shows the result of this analysis for the IFS and IRDIS. The detection limit corresponds to the limit where fake planets are detected at an SNR of 5 or higher. The penalty term induced by small sample statistics is taken into account \citep{mawet2014}. The IFS detection limit shows excellent performance, with a contrast of 14.30~mag reached at 0.2\as and 16.3~mag at 0.4\as. There is a slight increase of the limit around 0.6\as, which corresponds to the signature of the AO-cutoff frequency averaged over all wavelengths, as all wavelengths are combined together for this analysis. This represents the best contrast ever reached in high-contrast imaging at these angular separations.

For IRDIS, the fact that only two wavelengths are available significantly affects the performance at small separation due to the spectral self-subtraction effects. There is a steep rise of the limit for separations $\lesssim$0.5\as, but the limit joins that of the IFS at 0.6\as, and it remains rather flat around 16.7~mag in the 0.8--1.1\as range. Similarly to the IFS, the AO-cutoff frequency in $K$~band creates a slight increase around 1.1\as. Then the increase in sensitivity follows nicely the shape of the halo to reach a contrast of 19~mag at 3.7\as. In Fig.~\ref{fig:detection_limit_mag} we compare our SPHERE limits with the best limits on Sirius~A published by \citet{thalmann2011}, which were obtained in saturated imaging with Subaru/IRCS in the Br~$\alpha$ narrow-band filter (4.05~\mic). In the speckle-limited regime of these observations ($\lesssim$3\as), the gain of extreme AO and an efficient coronagraph becomes evident, with a gain of 3.5~mag at 1\as, 4.3~mag at 0.7\as and access to separations as small as 0.2\as. We would likely have accessed even smaller regions if the ADCs had worked properly. Nonetheless, these results show the huge potential of high-contrast imaging with the new generation of instruments.

We also compare our SDI+ADI analysis with an ADI-only analysis for both the IFS and IRDIS. For the IFS, we bin the data spectrally to an $H$ broad-band equivalent before performing the ADI analysis, and for IRDIS we similarly add together the data of the K1 and K2 filters. The SDI+ADI performance for the IFS shows a clear gain of almost 2~mag at 0.2\as and $\sim$1.5~mag beyond 0.4\as with respect to an ADI-only analysis. There is no clear evidence that SDI is failing at these separations thanks to the very large bandpass of the IFS in the \texttt{IRDIFS\_EXT} mode (52\%), but the effect would surely become much stronger at smaller separations where we would get close to the bifurcation radius (148~mas, see Sect.~\ref{sec:adi+sdi}). Another important point to keep in mind is that the use of PCA over temporal and spectral dimensions mitigates the adverse effect that SDI can have at small separations, especially if the conditions are stable in time and the amount of field rotation is large. For IRDIS, we see that ADI starts to give a slightly improved sensitivity at 0.3\as and below, but inside 0.3--2.0\as the gain of using SDI is clearly visible. At larger separations the gain is only marginal.

\subsection{Physical detection limits using fake planets with realistic spectra}
\label{sec:simulated_spectra}

The previous approach is useful to obtain an idea of the raw instrument performance in terms of contrast, but it is not realistic in terms of defining what physical objects can actually be detected. Indeed, the spectra of cool planetary mass objects are very far from flat, presenting strong molecular absorption features in the near-infrared when \Teff decreases \citep[e.g.][]{allard1997,burrows2006}. Atmospheric model grids coupled with evolutionary models \citep{burrows1997,chabrier2000,baraffe2003} have been commonly used in imaging surveys or for individual targets to convert single-band contrast limits into a mass limit a posteriori \citep[e.g.][]{vigan2012,meshkat2013,chauvin2015}. However, multi-wavelength data obtained with an IFS or a dual-band imager adds a level of complexity. Spectral features have a strong impact on the detectability of these objects and the spectral self-subtraction cannot be calibrated easily when a large number of spectral channels are involved.

\begin{figure}
  \centering
  \includegraphics[width=\columnwidth]{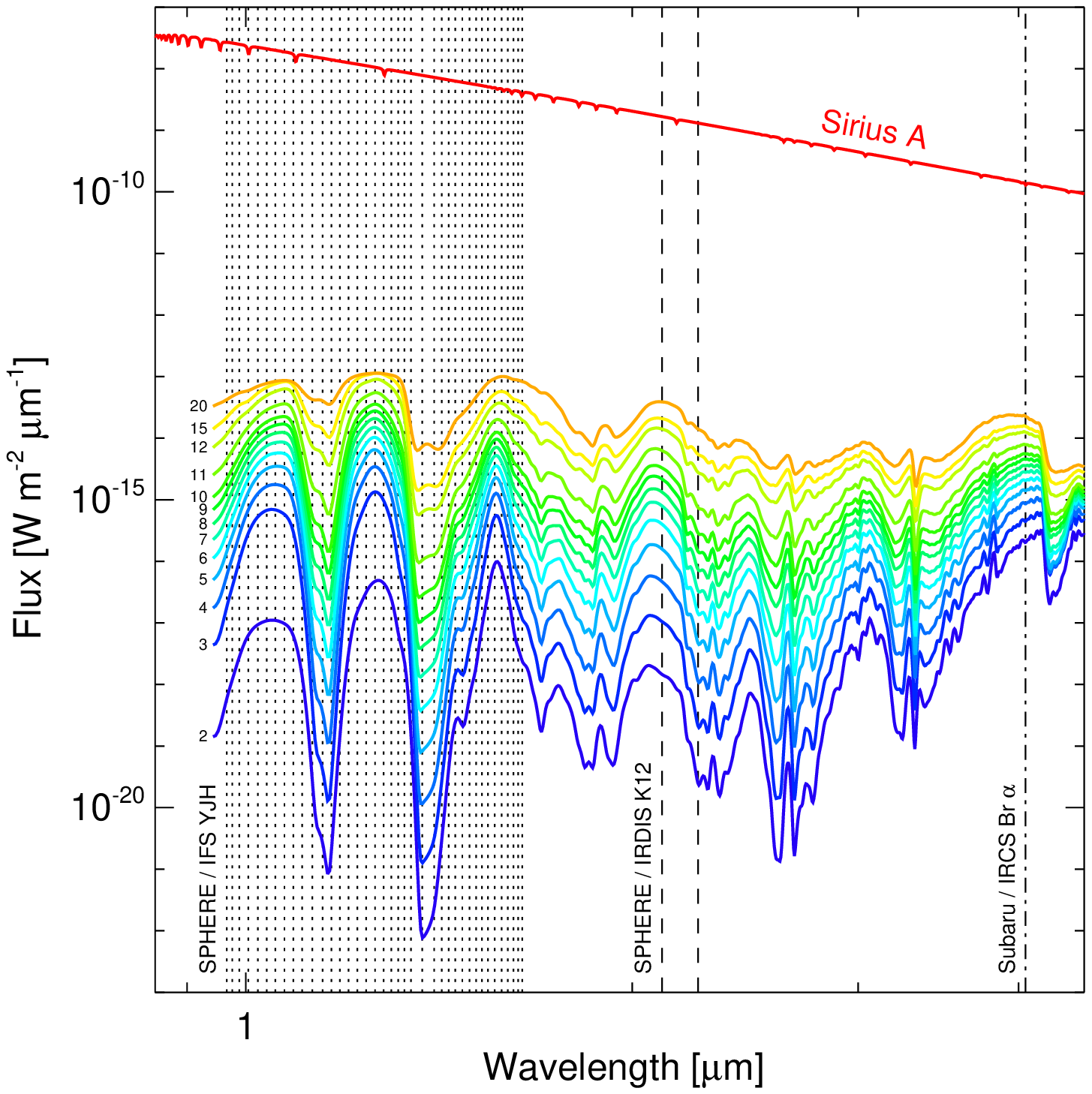}
  \caption{Comparison of the Sirius SED \citep{bohlin2014} with BT-Settl atmospheric models \citep{allard2013,baraffe2015} calculated for increasing masses from 2 to 20~\MJup using the \citet{baraffe2003} evolutionary tracks. The dotted, dashed and dot--dashed vertical lines show the central wavelengths of the IFS spectral channels, IRDIS K1 and K2 DBI filters and Subraru/IRCS Br~$\alpha$ filter respectively. The expected contrast for each planet is calculated per-channel by integrating numerically the flux of the planet model and dividing it by the flux of Sirius~A.}
  \label{fig:sirius_models}
\end{figure}

For our analysis, we make use of the most recent theoretical developments by using the BT-Settl atmospheric models \citep{allard2013,baraffe2015}, linked to the evolutionary models for the AMES-Cond grid \citep{baraffe2003}. The BT-Settl model couples a cloud model to the 1D general stellar atmosphere code \texttt{PHOENIX} \citep{allard1994,allard2001,hauschildt1997}. The model considers the formation of a cloud deck composed of up to more than 200 condensate species, with optical data for 50 species included in the radiative transfer. The grain size and density, and the abundances of chemicals in the gas phase, including the effect of element depletion induced by the grain formation, are computed layer per layer through the photosphere following a comparison of the time-scales for nucleation, condensation, gravitational settling or sedimentation, and mixing. Once rained out, the grains and their constituent elements are removed from the composition of that layer and all higher levels of the atmosphere. These models can predict the flux at the surface of a given object only defined by (\logg, \Teff), and the assumed (solar) composition. For the masses probed here, below $\sim$15~\MJup the models predict the settling of most of the primary iron and silicate clouds below the photosphere. For lower \Teff, corresponding to 5--10~\MJup, the formation of a secondary cloud layer within the photosphere, sitting above the primary cloud layer and consisting of MnS and Na$_2$S, then of KCl, NaCl and some ZnS, is predicted, confirming the studies of \citet{morley2012}. In the 2014 release of the models used here, opacities for all the above revised alkali cloud species have been added, but a specific grid is calculated to cover a lower domain of \Teff because of the age of our target. This extends into the regime of water ice and ammonium hydrosulphide formation tentatively confirmed in the latest observed Y dwarfs \citep{beamin2014}.

The spectra are simulated on a grid of \Teff and \logg, covering the range 200--1200~K and 3.5--4.5~dex respectively. We use evolutionary tracks calculated for the upper limit age of Sirius, 250~Myr \citep{liebert2005}, to calculate the expected \Teff, \logg and radius for planets of mass from 2 to 15~\MJup by steps of 1~\MJup, and from 15 to 25~\MJup by steps of 2.5~\MJup. Then, for each wavelength from 0.9 to 2.3~\mic, we interpolate into the (\Teff,\logg) grid of synthetic spectra to produce the spectrum expected for each mass. Finally, to obtain the flux received on Earth (at the top of the atmosphere) from a planet orbiting Sirius, the spectrum is multiplied by a dilution factor equal to $(R_{\mathrm{p}}/d)^2$, where $R_{\mathrm{p}}$ is the planet radius determined from the evolutionary tracks and $d = 2.637$~pc is the distance to Sirius. Then, the spectrum of each planet is numerically integrated in the IFS spectral channels and the two IRDIS filters to get an estimation of the integrated flux. To calculate the contrast at which the fake planets must be injected in each spectral channel, we use high-quality spectral energy distribution (SED) of Sirius~A published by \citet{bohlin2014}. The SED is also numerically integrated in each channel, and the contrast is simply obtained from the ratio of the planet flux with the Sirius flux. The Sirius SED and model spectra calculated for different masses are shown in Fig.~\ref{fig:sirius_models}. From this analysis, we expect contrasts in $J$~band of 15.5~mag and 13.7~mag for 5 and 10~\MJup planets respectively, and 17.5~mag and 14.3~mag for the same masses in $K$~band.

\begin{figure}
  \centering
  \includegraphics[width=\columnwidth]{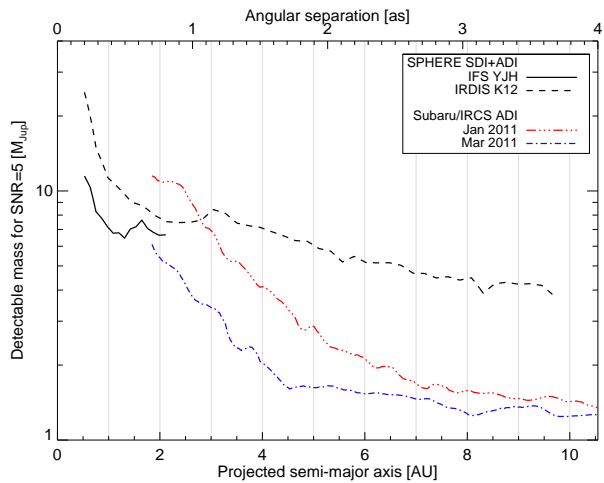}
  \caption{SPHERE sensitivity in mass to obtain a detection at SNR=5 with the IFS (plain curve) and IRDIS (dashed curve), based on the injection of fake planets with spectra based on the BT-Settl synthetic spectra and evolutionary tracks (see Section~\ref{sec:simulated_spectra} for details). These limits are compared to the contrast limits of \citet{thalmann2011} in the Br~$\alpha$ filter at 4.05~\mic that we converted into mass using the same set of models as for our SPHERE analysis. We discuss the comparison between the Subaru/IRCS and SPHERE/IRDIFS limits in Section~\ref{sec:discussion}}
  \label{fig:detection_limit_mass}
\end{figure}

To estimate the detection limits, the process detailed in Sections~\ref{sec:detlim_method_description} and \ref{sec:flat_spectrum} is repeated. Fig.~\ref{fig:detection_limit_mass} shows the final detection limits in physical units of projected orbital semi-major axis and mass of Jupiter. Here again the effect of small sample statistics is taken into account. Despite the extreme brightness of Sirius~A, we are able to reach the planetary-mass regime with both the IFS and IRDIS inside of 0.5\as (1.3~au). We are sensitive to masses below 10~\MJup for all projected semi-major axes outside of 0.7~au. Here again, the wide spectral coverage of the IFS shows its strength by allowing the detection of masses below 12~\MJup within the 0.5--2.1~au range. In fact, the IFS detection limit is even better than IRDIS at the edge of its FoV. Similarly to the stellar-like spectrum cased detailed Section~\ref{sec:flat_spectrum}, the IRDIS limit gets worse towards very small separations because of self-subtraction effects. 

\subsection{Comparison to a simplified approach}

\begin{figure}
  \centering
  \includegraphics[width=\columnwidth]{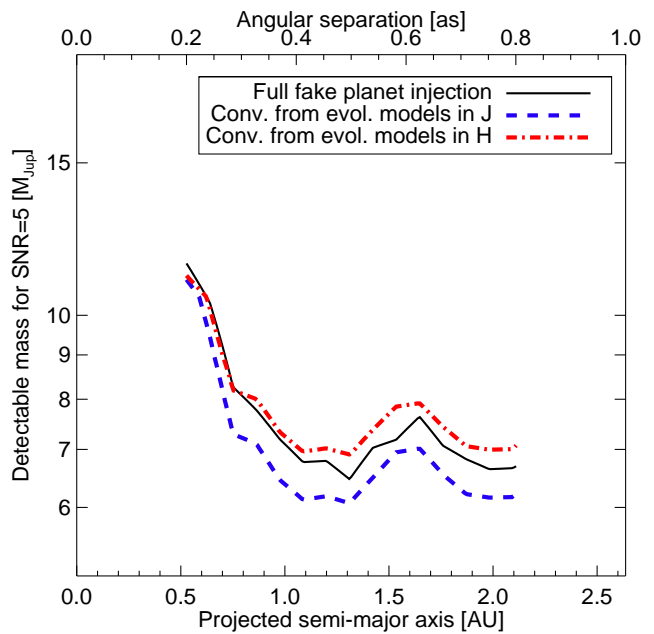}
  \caption{Comparison of the IFS detection limits obtained using our approach, based on the injection and redetection of fake planets with a realistic spectra, and using the conversion of the contrast limit from Fig.~\ref{fig:detection_limit_mag} using the evolutionary models in $J$~band or in $H$~band, as proposed in \citet{mesa2015} and Zurlo et al. (submitted).}
  \label{fig:detection_limits_comparison}
\end{figure}

When the stellar SED is readily available or easy to determine by model-fitting (e.g. Vigan et al. submitted), the approach we have described in the previous sections is extremely straightforward to implement. In particular, it has the advantage of solving one of the usual biases related to ADI and SDI analysis, which is the estimation of the self-subtraction effects that are not always easy to quantify, especially when considering multi-wavelength data. Indeed, contrary to the a posteriori conversion of detection limits usually employed up to now, our approach embeds the model-dependent evolutionary part directly into the analysis of the data. At the end of the analysis, the detection level of the injected fake planets provides the mass detection limit directly; therefore, the flux losses do not need to be determined. The data need to be fully reprocessed many times to obtain a robust estimation of the detection limit which makes this process very computationally demanding.

We use the Sirius data set to compare our approach with the less demanding one detailed in \citet{mesa2015} and Zurlo et al. (submitted) for the estimation of the detection limit in physical unit. In their approach, the data analysis is also based on PCA, but the detection limit in contrast is estimated from the residual noise in the final images, corrected for the self-subtraction effects using injection of fake planets close to the noise floor, and normalized to the median peak flux of the PSF at all wavelengths. Then, this detection limit is converted into physical units using evolutionary models calculated in either the $J$- or $H$~band broad-band filters. This approach provides faster results, as it is only necessary to fully process the data twice, however, it does not take into account all of the spectral information available in the IFS data. This information is lost (at least partialy) when converting the final contrast detection limit using only predictions in a broad-band filter.

The results of our comparison are showed in Fig.~\ref{fig:detection_limits_comparison}. In this figure we compare our IFS mass detection limit from Fig.~\ref{fig:detection_limit_mass} to the contrast limit from Fig.~\ref{fig:detection_limit_mag} that we converted into mass using the same set of models. To obtain the limit, the expected contrast between Sirius~A and planets at increasing masses is calculated in broad-band $J$ and $H$ filters. Then, the final mass limit is obtained by interpolating this expected contrast at the values defined by the contrast limit. This approach shows very good agreement for the conversion using the $H$~band, with differences of 5\% at most with respect to the limit derived using the injection of fake planets. In $J$~band the conversion is less accurate, with differences of 10\% on average. 

The good agreement between our method and the conversion using the evolutionary models in the $H$~band could be anticipated because the contrast is expected to be smaller in the $H$~band: if a planet is detected in the IFS data, most of the strength of its signal will come from the $H$~band; so converting the contrast detection limit using luminosity predictions in the $H$~band will likely lead to a good approximation of the mass limit. Still, this comparison is an important step for on-going large direct imaging surveys, as it shows that mass limits could be derived from simple contrast curves rather than going into much more advanced and time-consuming analyses. However, we note that one specific case does not make a rule. A more systematic study taking into account stars of different spectral types (and thus colours) and planetary atmosphere models would be required to draw more general conclusions.

\section{Discussion}
\label{sec:discussion}

\begin{figure}
  \centering
  \includegraphics[width=\columnwidth]{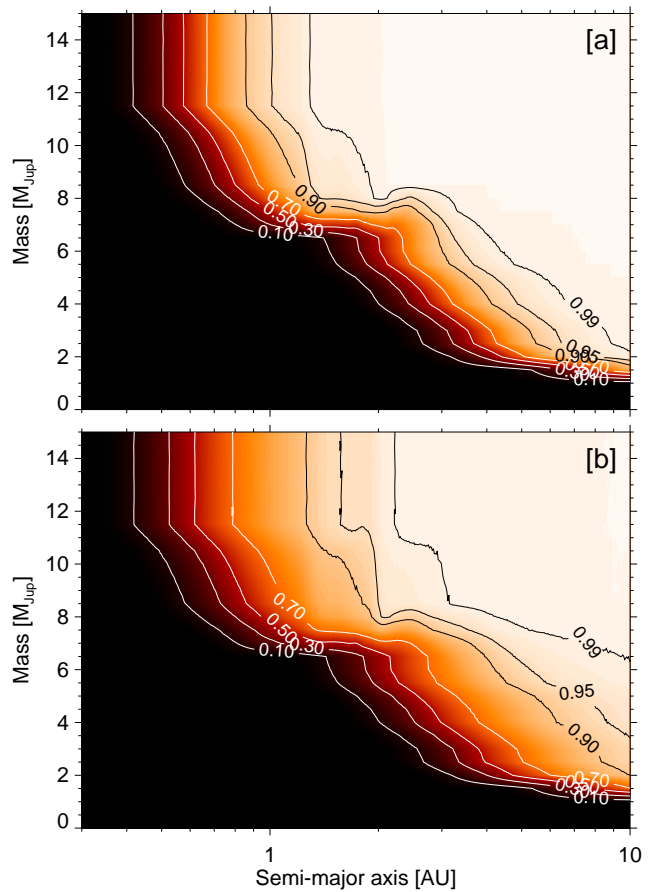}
  \caption{Mean probability of at least one detection of a substellar companion around Sirius~A in the combined Subaru/IRCS and SPHERE/IRDIFS data sets as a function of the companion mass and semi-major axis. We consider [a] the case of coplanar orbits with the Sirius~A--B system ($i=136.6\deg$) or [b] the case of unconstrained inclination $i$. Compared to the similar study by \citet{thalmann2011}, we are able to place constraints down to 1~au for giant planets more massive than 7--8~\MJup thanks to our new SPHERE/IRDIFS observations, which offer sensitivity to planetary-mass objects down to 0.2\as (0.5~au).}
  \label{fig:mc_analysis}
\end{figure}

In Section~\ref{sec:determination_detection_limits} we demonstrated that the new VLT/SPHERE instrument keeps its promises in terms of reachable contrast in the near-infrared for very bright stars. Fig.~\ref{fig:detection_limit_mag} illustrates the new parameter space, in angular resolution and contrast, that the new high-contrast imagers will access compared to the previous generation instruments.

In the case of Sirius, the brightest star in the sky, this high-contrast at small separation translates to being sensitive to planets below 10~\MJup at separations of less than 1~au at its nominal age of 250~Myr. However, it is known from previous studies \citep{rameau2013,skemer2014} that $L$~band imaging is even more favourable for young giant exoplanets detection because of the decreased stellar flux and increased planet flux at longer wavelengths. This is illustrated in Fig.~\ref{fig:sirius_models}, where we see that the Sirius flux drops by a factor $\sim$100 between $J$~band and the Br~$\alpha$ filter at 4.05~\mic, and by a factor $\sim$10 between $K$~band and this filter. In this same range, the peak flux of low mass planets also increases by factors of 10 or more, resulting with expected contrasts of only 12.4 and 11.0~mag for 5 and 10~\MJup planets at 4.05~\mic respectively.

For comparison of our IRDIFS detection limits to the Subaru/IRCS limits from \citet{thalmann2011} in the Br~$\alpha$ filter, we converted their contrast limits into mass using the same set of models. The use of this more recent grid of models has the effect of improving slightly their mass limits with respect to their published ones, which were obtained using the COND evolutionary models \citep{baraffe2003}. The reader should however keep in mind that predictions in the 1--3~\MJup range are still very uncertain, whatever the type of model used. Still, using more advanced models is perfectly justified despite the uncertainties, which equally apply to the COND atmosphere models. This comparison of limits is particularly interesting, as it clearly shows the usefulness of having access to longer wavelengths for planets detection. Indeed, Subaru/IRCS allows to reach lower masses than SPHERE: at 0.7\as (1.8~au), they are sensitive to planets with masses as low as 6~\MJup (against $\sim$7~\MJup with SPHERE), and they reach 1.5~\MJup at 1.5\as (4~au). 

Despite these impressive limits in $L$~band, one has to keep in mind the limitations of observing at long wavelengths: (i) the final sensitivity is usually dominated by the sky background noise except in a few very bright cases like Sirius~A, and (2) the inner-working angle is also usually wider because of the larger wavelength, although new devices such as the annular groove phase mask \citep{mawet2005} can provide access to inner-working angles almost down to the diffraction limit of 8~m telescopes \citep{absil2013}. Nonetheless, our comparison shows the exemplary complementarity between new high-contrast imagers in the near-infrared and other imaging instruments in the $L$- or $M$~band (e.g. NACO at the VLT, NIRC2 on Keck).

We use this complementarity to place constraints on the presence of additional companions around Sirius~A. For this purpose, we perform orbital Monte Carlo simulations where planets are simulated on a grid of mass and semi-major axes, in the range 1--30~\MJup and 0.2--100~au respectively. At each point of the grid, $10^{5}$ orbits are simulated with random eccentricity, longitude of periastron, longitude of ascending node and time of passage at periastron. For the inclination, we simulate two cases: either a random inclination (statistical weight proportional to $\sin i$) or a fixed inclination equal to $136.6 \deg$, which corresponds to a co-planar case with the Sirius~A--B system. For each simulated orbit, the expected projected separation of the planet is calculated for the three epochs where we have detection limits (two for IRCS, one for SPHERE), and we check the detectability of the planet with respect to the detection limits. For each point of the grid, the likelihood of detection is calculated as the fraction of planets detected at least one time over the total number of simulated planets. The results are presented in Fig.~\ref{fig:mc_analysis}.

Compared to the similar study by \citet{thalmann2011}, our new SPHERE/IRDIFS observations bring strong constraints on the existence of giant planets within the 0.6--2~au range where previous observations lacked good sensitivity. The co-planar case is better constrained towards small semi-major axes, and in such a case we are able to reject with 90\% probability the existence of a 10~\MJup planet at 1~au. In the more general case this probability decreases to 70\%, which is still high. At smaller separations or lower masses, the completeness values drop significantly and we cannot draw any firm conclusions. This is nonetheless the first time that we can probe with high confidence the region below the critical semi-major $a_{\mathrm{c}} = 2.17$~au where the simulations by \citet{holman1999} predict that stable orbits can be found for planetary companions. Of course, we cannot reject the possibility of the existence of planets that could not be detected, either because their mass is too low to be detected by imaging or because they are currently hidden behind or too close to the star. At small semi-major axes, the probability of the latter increases significantly and multiple epochs would be necessary to rule out giant planets at 1~au or below at high confidence.

Finally, we conclude by noting that because Sirius~A is a relatively massive young star of type A1V, it is not adapted to the search of low-mass companions by the radial velocity methods, so imaging remains one of the most viable options to detect a planet in this system.

\section{Conclusions}
\label{sec:conclusions}

The Sirius system is certainly among the most intriguing stellar systems at close proximity to Earth. We have used high-contrast imaging in the near-infrared with the new SPHERE instrument, recently commissioned at the VLT, to get the closest ever look at Sirius~A. This new generation instrument combines extreme adaptive optics and modern coronagraphs to allow the detection of very faint planetary companions around nearby, young stars. 

As for any new instrument, especially when relatively complex like SPHERE, the data reduction of the first data sets by the community can be complicated. We have provided some useful guidelines on how to improve basic calibration steps for the IFS, as well as how to deal with problems that other users could encounter in the future such as saturated off-axis PSF and inaccurate or unstable centring of the star behind the coronagraph during the observations. Thanks to these improved calibration steps, we were able to get the most of this high-quality data set.

In terms of raw contrast, we present what is the best contrast ever obtained in the 0.2\as--0.5\as range with a direct imaging instrument. Using fake planets with a stellar-like spectrum, we demonstrate that we reach with the IFS a contrast of 14.3~mag at 0.2\as and 16.3~mag at 0.4\as. With IRDIS, the self subtraction effects limits our sensitivity at small separations ($\lesssim 0.7\as$), but we demonstrate the very nice complementarity between the two SPHERE near-infrared instruments: IRDIS reaches the same sensitivity as the IFS at the edge of the outer IFS FoV, providing a perfect wider-field complement up to 4\as where it reaches a contrast of $\sim$19.0~mag. This exceptional detection limit likely represents SPHERE at its very best because of the very high brightness of the guide star ($R = -1.46$), the close to perfect observing conditions ($\sim$0.4\as seeing) and very large amount of field rotation ($106\deg$).

Despite their exceptional sensitivity, our observations bring another non-detection of companions around Sirius~A. Multiple analyses of the IRDIFS data using approaches based on ADI or SDI+ADI do not reveal any reliable candidate. Starting from this non-result, we presented a new approach for the computation of the IFS detection limits based on the use of the stellar SED, evolutionary models and associated atmospheric models. While recent high-contrast instruments are all equiped with an IFS, the problem of deriving reliable detection limits for this type of high-contrast data has been mostly overlooked so far. However, it constitutes a necessary step for the large-scale direct imaging surveys that are currently on-going with SPHERE or its North-American equivalent GPI. In this context, we compared our approach to a more simple one where the contrast detection limit is directly converted to mass using evolutionary predictions in the $H$ broad-band filter. The two approach are in good agreement in the Sirius case, however a more detailed study would be required to draw firm conclusions.

Our limits around Sirius~A show that we are sensitive to giant planets more massive than 11~\MJup at 0.5~au, 6--7~\MJup in the 1--2~au range, and down to $\sim$4~\MJup at 10~au. Outside of this range, previous limits obtained by \citet{thalmann2011} with Subaru/IRCS in the Br~$\alpha$ filter at 4.05~\mic are more sensitive thanks to the decreased stellar flux and increased planet flux at longer wavelengths. This illustrates the complementarity that exists between new near-infrared high-contrast imagers and $L$- or $M$~band AO imagers for the search of planetary-mass companions.

Monte Carlo simulations based on the SPHERE/IRDIFS and Subaru/IRCS detection limits show that our observations provide for the first time meaningful constraints on giant planets around Sirius~A in the 0.5--1.5~au range, with about 50\% probability for the existence of an 8~\MJup planet at 1~au. A second epoch observation with SPHERE would probably strengthen these conclusions because at small semi-major axes the probability increases that the planet is behind or too close to the star. In any case, additional SPHERE data will not bring better constraints for planets less massive than 7--8~\MJup or at separations smaller than 0.5~au, so there is still plenty of room for smaller planets that could not be detected with direct imaging. The Sirius system will keep its cloud of mystery for the foreseeable future.

\section*{Acknowledgements}

AV and GS acknowledge support in France from the French National Research Agency (ANR) through project grant ANR10-BLANC0504-01. DM acknowledges support from the ``Progetti Premiali'' funding scheme of the Italian Ministry of Education, University, and Research. DH acknowledges support from the European Research Council under the European Community's Seventh Framework Programme (FP7/2007-2013 Grant Agreement no. 247060) and from the Collaborative Research Centre SFB 881 `The Milky Way System' (subproject A4) of the German Research Foundation (DFG).

Model atmospheres have been computed at the {\sl P{\^o}le Scientifique de Mod{\'e}lisation Num{\'e}rique} of the {\'E}cole Normale Sup{\'e}rieure de Lyon and at the {\sl Gesellschaft f{\"u}r Wissenschaftliche Datenverarbeitung G{\"o}ttingen} in collaboration with the Institut f{\"u}r Astrophysik G{\"o}ttingen.

The authors are extremely grateful to C. Thalmann for kindly providing his published Subaru/IRCS detection limits. They would also like to thank the ESO observers that made their best effort to acquire the data during the SPHERE science verification, and D. Mawet and J. Milli for their explanations on the small sample statistics. Finally, they would like to thank L. Mugnier and N. V\'edrenne, from ONERA, for the spatial rescaling algorithm detailed in Appendix~\ref{sec:spatial_scaling_algorithm}.

\bibliographystyle{mnras}
\bibliography{paper}

\appendix

\section{Improving the SPHERE/IRDIFS data reduction}
\label{sec:improving_irdifs_data_reduction}

\subsection{SPHERE/IFS pre-processing pipeline}

Hoping that it will help the community to work with SPHERE/IFS data, we make our pre-processing pipeline available at the following address: \\

\url{http://people.lam.fr/vigan.arthur/} \\

\noindent It contains a set of tools that simplifies the process to go from the raw data to the properly calibrated $(x,y,\lambda)$ data cubes, providing a good starting point for subsequent SDI and ADI analyses.

\subsection{Correction of the IFS wavelength calibration}
\label{sec:correction_ifs_wavelength_calibration}

An important calibration for the IFS is the determination of the wavelength of each of the spectral channels in the data cubes. The DRH provides a good estimation of the wavelength of each channel, but one can easily realize that the accuracy is not sufficient by performing a spatial rescaling of each channel by a factor equal to $\lambda_{0}/\lambda_{i}$, where $\lambda_{0}$ and $\lambda_{i}$ are respectively the wavelength of the first and the $i$th spectral channels. After such a rescaling, all the diffraction features (AO control radius, halo, speckles) should remain stable when going through all the spectral channels. However, when using the wavelength estimation from the DRH, the diffraction features are not completely stable after rescaling, which indicates that the wavelength estimation is imperfect. To improve the wavelength estimation, we perform a new calibration in two steps that allow (1) to determine the relative scaling factor between the spectral channels, and (2) determine a reference channels for which the absolute wavelength is known precisely.

\begin{figure}
  \centering
  \includegraphics[width=\columnwidth]{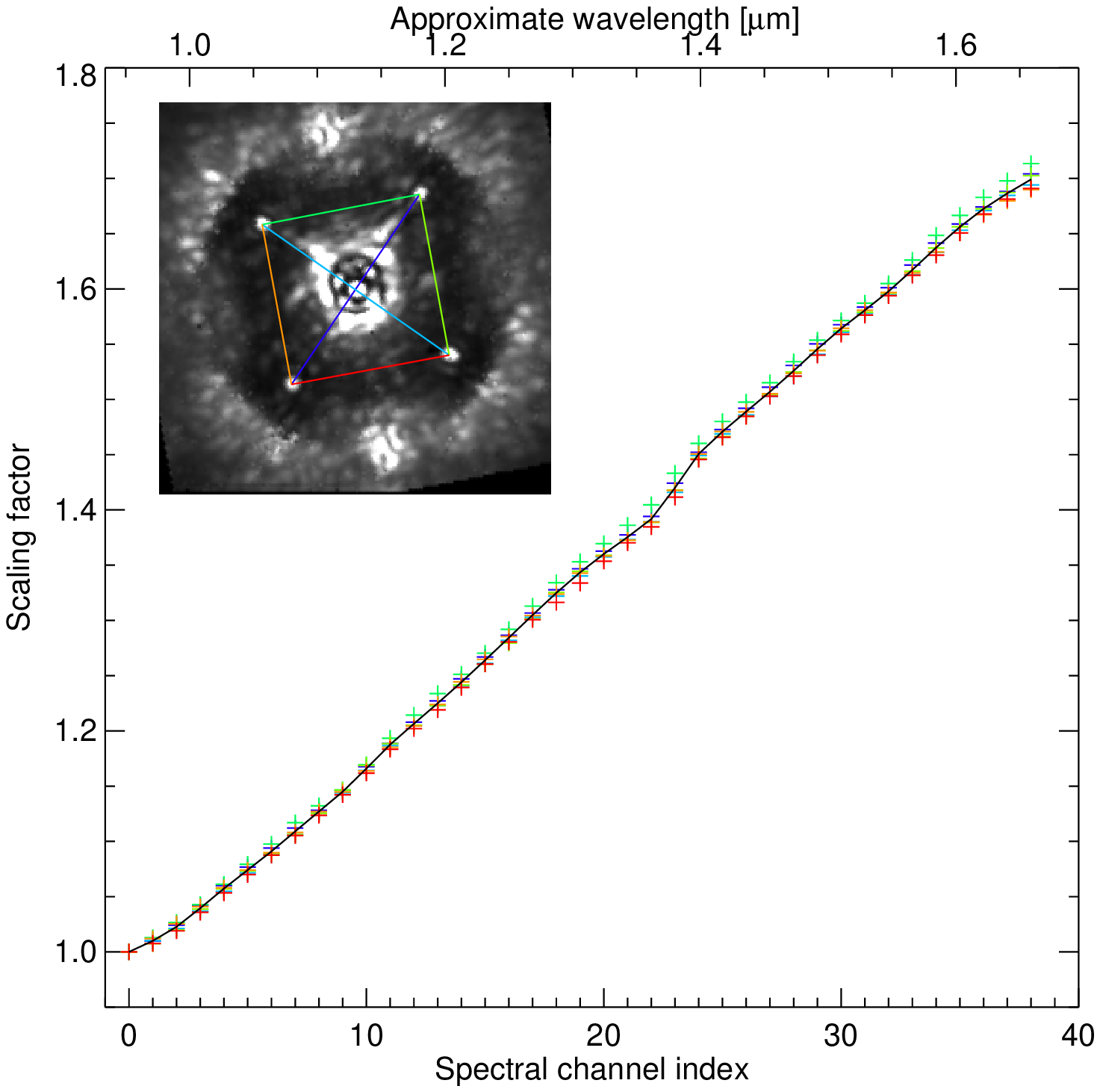}
  \caption{Scaling factor with respect to the first spectral channel as a function of the spectral channel index. The top axis gives the approximate wavelength of the channels. In each channel, six distances are measured using the satellite spots to determine the average scaling factor (black curve). The inset image shows the star centre frame at a wavelength of 1.24~\mic with the four spots and the six measured distances in different colours.}
  \label{fig:ifs_scaling_factor}
\end{figure}

For step (1), we use the star centre frames described in Section~\ref{sec:observations}. In these frames, the periodic modulation introduced on the deformable mirror creates replicas of the stellar PSF that appear as four symmetrical satellite spots at a separation of $\sim$14~\lsd from the star. The spacing of the spots is proportional to wavelength, so by measuring the spacing for channel $i$ with respect to channel 0, we are able to determine accurately the wavelength scaling factor. To increase the accuracy, we measure six different distances between the four spots (two between opposite spots, four as the sides of the squares). The final scaling factor is taken as the average scaling factor determined for the six distances, as illustrated in Fig.~\ref{fig:ifs_scaling_factor}. There are small differences in the scaling factors determined independently, and from the departure between the average scaling factor and the independent factors, we estimate an accuracy better than 0.93\% (0.26\% on average) on the determination of the scaling factor. 

\begin{figure}
  \centering
  \includegraphics[width=\columnwidth]{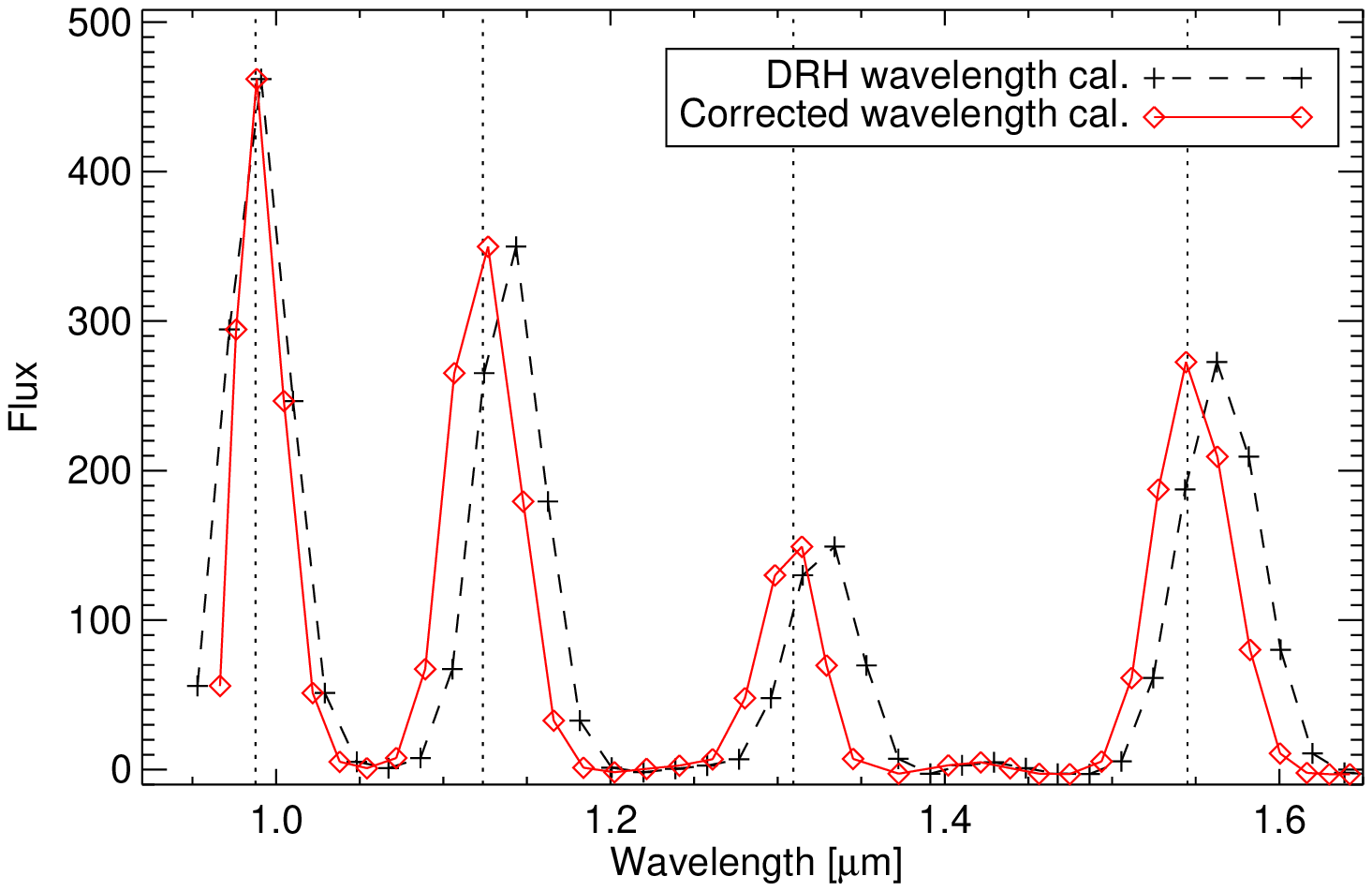}
  \caption{Median flux measured in the wavelength calibration processed as science data as a function of the DRH wavelength (dashed black curve) and after correction (red plain curve). The dashed vertical lines show the wavelength of the laser lines that are used for the wavelength calibration of the IFS.}
  \label{fig:ifs_wave_cal}
\end{figure}

The scaling factor is only a relative measurement that does not provide information on the absolute value of the wavelength of each spectral channel. To improve the absolute wavelength calibration, we need to determine the wavelength of one reference spectral channel, and the wavelengths of all the others will then be extrapolated using the scaling factor. For the wavelength calibration, the detector observes a flat source illuminated by four laser lines through the IFU, creating four spots with precisely known wavelengths for each lenslet. In the DRH recipe, for each lenslet, the position of these four spots is measured and a third order polynomial is fitted to determine the wavelength of all pixels corresponding to the lenslet. For the step (2) of our wavelength calibration refinement, we make use of the wavelength calibration raw frame, which we process in exactly the same manner as the science data. The resulting $(x,y,\lambda)$ data cube shows flat images with flux variations as a function of wavelength, which correspond to the presence or not of laser lines in the original raw frame. The median flux in each of the spectral channels of this cube is plotted as a function of the DRH-determined wavelength in Fig.~\ref{fig:ifs_wave_cal}. The 4 laser lines are clearly visible, but the peaks appear shifted with respect to the wavelength of the lasers, indicating some error or offset in the DRH wavelength calibration. To correct for this, we implemented a Levenberg-Marquardt least-squares minimization routine in which we perform the following steps:

\begin{itemize}
  \item We make a Gaussian fit of the position of the laser peaks as a function of the spectral channel index.
  \item We attribute a wavelength to one of the channels (reference channel) and we extrapolate the wavelength of all the other channels using the scaling factor discussed previously. For the reference channel we take as first guess the wavelength determined from the DRH.
  \item We determine the expected wavelengths at the position of the peaks determined in the first step by interpolating (linearly) the newly defined wavelength calibration.
  \item Finally, we measure the difference between these expected wavelengths and the known wavelengths of the lasers.
\end{itemize}

\noindent The only free parameter is the wavelength of the reference channel, and the minimization is performed on the maximum of the difference (in absolute value) calculated in the last step. The median flux as a function of the corrected wavelength is also plotted for comparison in Fig.~\ref{fig:ifs_wave_cal}. This time the peaks properly match the expected wavelengths, and the difference is between 0.37 and 0.85~nm, while it was between 1.7 and 19.4~nm with the DRH wavelength calibration. Overall, combining the errors on the determination of the scaling factor and the wavelength of the reference channel, we estimate an error on the wavelength calibration of the order of 1.4~nm at most.

Our procedure assumes that the position of the spectra on the IFS detector, and hence the wavelength calibration, is stable between the time when the science data is taken and the time when the wavelength is calibrated. SPHERE is located on one of the Nasmyth platforms of VLT Unit Telescope 3, which is mechanically very stable. The main temporal evolutions of the calibrations that are observed in SPHERE are due to bench temperature variations. For the IFS, the position of the spectra evolves by $\sim$0.3~pixel/deg (internal ESO monitoring). During the Sirius observations, the temperature of the IFS was stable at 15.9$^\circ$C, and for the wavelength calibration taken in the morning the temperature was 15.6$^\circ$C, resulting in a shift of 0.09~pixel of the spectra. We consider this variation to be negligible and do not attempt additional corrections for this effect.

\subsection{Scaling of the off-axis proxy PSFs}
\label{sec:scaling_off-axis_proxy_psf}

\begin{figure}
  \centering
  \includegraphics[width=\columnwidth]{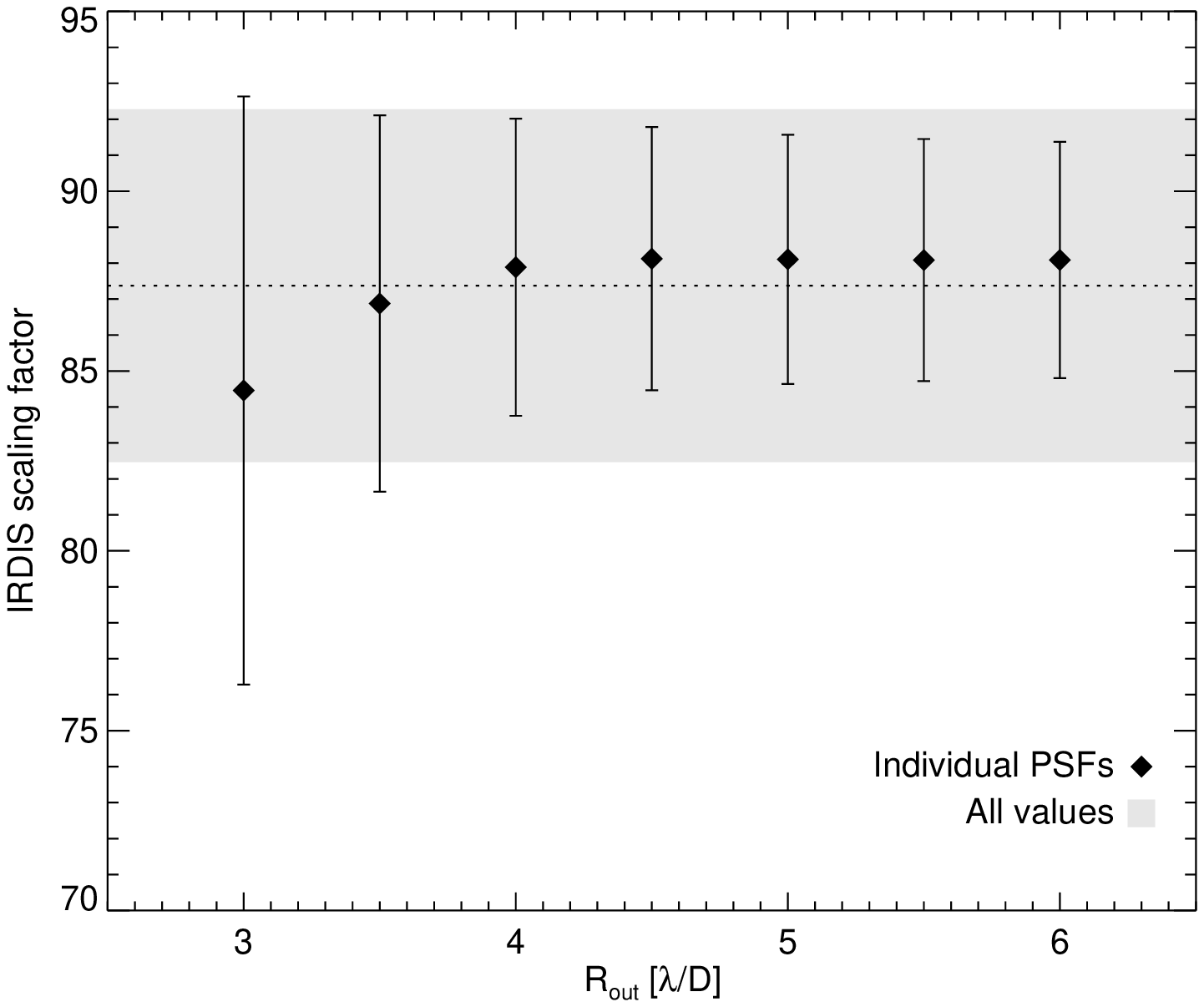}
  \caption{Scaling factor necessary for the $\beta$~Pictoris proxy PSF to match the photometry of Sirius~A in IRDIS the K12 filter pair. The factor has been determined by fitting the proxy PSF on individual DITs of the Sirius PSF (black diamond and associated error bars). The dotted line corresponds to the average of all values, and the grey area to the average plus or minus one standard deviation.}
  \label{fig:ird_scaling}
\end{figure}

\begin{figure*}
  \centering
  \includegraphics[width=\linewidth]{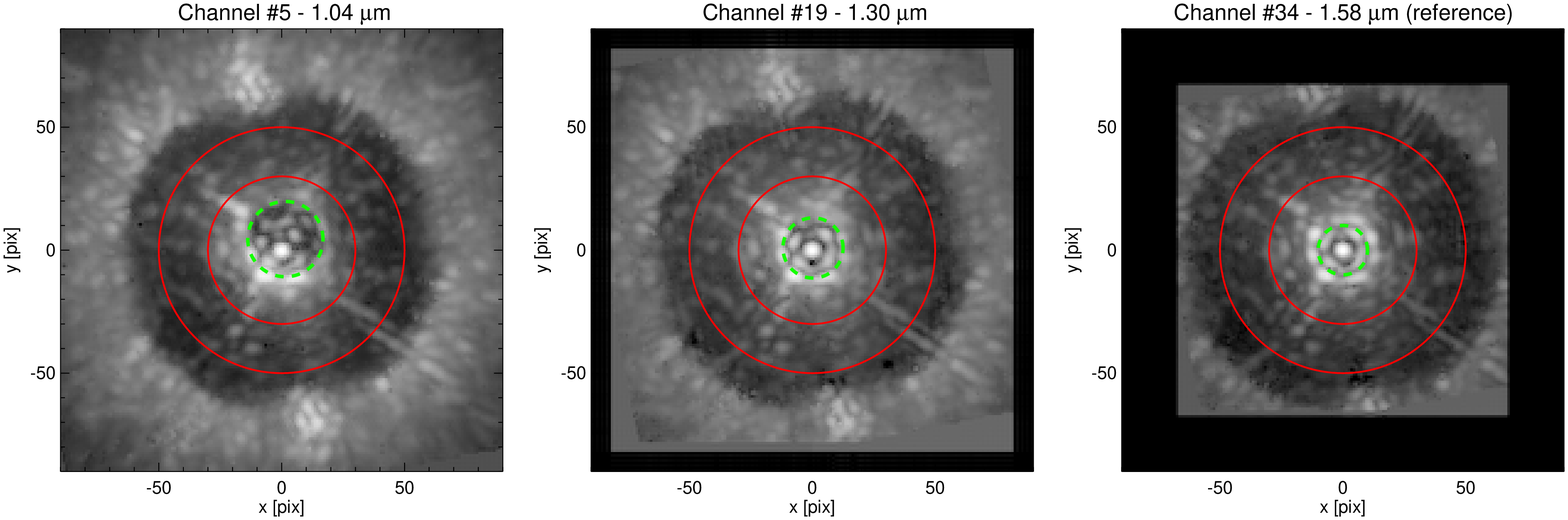}
  \caption{Spatially rescaled images in three spectral channels of the IFS taken from the same data cube. The red circles delimit the annulus where the residuals of the difference between the reference frame and the others are minimized in the centring procedure (see text for details). The green dashed circle shows the position and size of the coronagraphic mask. Each frame is centred with respect to the star. Because of the ADC tracking problem, the star centre and the coronagraphic mask centre do not correspond at all wavelengths. However, the speckle pattern and diffraction spikes of the telescope spiders match very well in all the channels. The optimization annulus is chosen to avoid the bright speckles at the edge of the coronagraph that vary a lot with wavelength because of the sub-optimal star centring, but also the aliasing at the border of the AO control radius.}
  \label{fig:ifs_centring}
\end{figure*}

As explained in Section~\ref{sec:off_axis_reference_psf}, the off-axis PSF was slightly saturated in the first five spectral channels of the IFS, and the core of the PSF in the K12 filters was saturated in IRDIS over a diameter of $\sim$\lsd. 

For the scaling the $\beta$~Pictoris proxy PSF, we base our analysis on the first Airy ring of both PSFs, which is well within the linearity range for Sirius with $\sim$3\,000 counts on average. To determine the scaling ratio, we use a fitting routine with three free parameters: the shift in $x$, the shift in $y$, and the flux scaling between the Sirius and proxy PSFs. During the fitting, the proxy PSF is shifted, scaled, and subtracted to the Sirius PSF. For IRDIS, we evaluate the standard deviation of the residuals in an annulus with an inner radius starting at 1.5~\lsd, and an outer radius $R_{\mathrm{out}}$ that we vary between 3 and 6~\lsd by steps of 0.5~\lsd. Figure~\ref{fig:ird_scaling} illustrates the average value and standard deviation of the scaling factor determined with the different values of $R_{\mathrm{out}}$ for IRDIS in the K1 filter. We perform this analysis on the three off-axis PSFs that were acquired for Sirius (30 NDIT for each). We adopt as the final scaling factor the average of all values determined for all values of $R_{\mathrm{out}}$, $87.4 \pm 4.9$. As a sanity check, we note that the difference of $K$~band magnitude between $\beta$~Pictoris and Sirius is equal to 4.83, corresponding to a factor of 85.5 that is perfectly consistent with our determination. The uncertainty on the determination of the scaling factor translates to a photometric uncertainty of 0.06~mag on the final PSF.

For the IFS, a similar analysis is done for the five first and four last spectral channels in which the PSF is just slightly saturated or out of the linear range. In these channels, the SNR of the Sirius PSF outside of the first Airy ring is not enough to extend $R_{\mathrm{out}}$ to more than 3~\lsd so we fit the proxy PSF only with this value. The estimated photometric uncertainty is of $\sim$0.2~mag for these channels. As an additional check, we compared the flux of the corrected off-axis PSF with the flux of the satellite spots in the star centre images, and they agree within the 0.2~mag uncertainty. The other channels are not saturated so no particular correction is applied to the PSF in these channels.

\subsection{Centring of the IFS coronagraphic data}
\label{sec:centring_ifs_coronagraphic_data}

In Section~\ref{sec:centring_coronagraphic_data} we described how the stopped near-infrared ADCs affect the centring of the star behind the coronagraphic mask, especially towards the shorter wavelengths in the IFS data. Fortunately, the DTTS ensures that the centring is accurate close to a wavelength of 1.5~\mic, where it measures the differential tip-tilt between the visible and the near-infrared. We verified that it is the case in our data using the star centre frames (see Section~\ref{sec:centring_coronagraphic_data}). We adopt as the reference wavelength the spectral channel number 35 ($\lambda_{35} = 1.58$~\mic), which means that we consider that the star is well centred on the coronagraph in this channel, and that it remains stable throughout the observation, despite the ADC problem. We use as reference centre in this channel the average of the three centres obtained from the three star centre frames. 

The second step of the analysis, for each IFS data cube, consists in using this reference channel to recentre all of the others. First, all channels are rescaled spatially using the wavelength-dependent scaling factor determined for the wavelength calibration in Appendix~\ref{sec:correction_ifs_wavelength_calibration}. The value of this scaling factor does not vary with time and can be determined once and for all for a given observation using the star centre frames. For an accurate spatial rescaling, we used an FFT-based procedure, detailed in Appendix~\ref{sec:spatial_scaling_algorithm}, which performs zero-padding both in the direct and Fourier spaces. After the rescaling, the speckle pattern has the same size in all frames, and we can find the shift between any spectral channel and the reference channel. In practice, the optimal shift is determined by minimizing the residuals in the difference between the reference channel and the other channels. The residuals are calculated in an annulus between 30 and 50 pixels. This area has been chosen because it is within the AO control radius in all channels but avoids the varying effect of aliasing at the edge of the control radius, and it contains bright speckles that can be easily matched between channels while avoiding the bright residuals close to the coronagraph edge. These residuals are particularly strong and varying because of the differential refraction that induces a decentring of the star behind the coronagraph with wavelength. Fig.~\ref{fig:ifs_centring} shows three spectral channels from the same cube with the optimization annulus, and the position and size of the coronagraphic mask. The frames are centred with respect to the star, resulting in a visible motion of the coronagraphic mask position as a function wavelength. However, the pattern of instrumental speckles is stable and most speckles within the optimization annulus can be seen in all three channels. 

\section{Spatial scaling algorithm details}
\label{sec:spatial_scaling_algorithm}

Accurate spatial rescaling of discrete images is very important for high-contrast imaging where techniques such as SDI \citep{racine1999} are heavily used for speckle noise attenuation. We detail below an FFT-based rescaling technique that makes use of zero-padding in both the direct and Fourier spaces for accurate rescaling of discrete data. 

For conciseness we consider one-dimensional signals, and the results generalize readily to two dimensions. Given $N$ samples of a signal $i$ sampled with a pitch $p_\alpha$, the quantity $\tilde{i}(k.p_\nu)$ given by:

\begin{equation}
  \label{eq-DFT}
  \tilde{i}(k.p_\nu) = \sum_{l=0}^{N-1} i(l.p_\alpha)\, e^{2j\pi\, l p_\alpha
  \, k p_\nu} ,
\end{equation}

\noindent is the discrete Fourier transform of $i$ at frequency $k.p_\nu$, where $p_\nu$ is the pitch in Fourier space. Because of the use of the fast Fourier transform (FFT), the pitch in Fourier and direct spaces are related by

\begin{equation}
  \label{eq-pitch-fft}
  p_\alpha p_\nu = 1/N,
\end{equation}

\noindent which means that the support $Np_\alpha$ is the inverse of the Fourier pitch.

If we modify the number of pixels in direct space by adding on each side of the table some null pixels \emph{before} performing the FFT, equation~\ref{eq-pitch-fft} becomes $p_\alpha p_{\nu'} = 1/N'$, where $N'$ is the new signal size after zero-padding in direct space, and $p_{\nu'}$ the new pitch in Fourier space.

If we proceed similarly in Fourier space by adding null pixels after performing the FFT, we finally obtain for the new spatial sampling in direct space after an inverse FFT:

\begin{equation}
  \label{eq-pitch-after-rescaling}
  p_{\alpha"} = (N'/N")  \, p_{\alpha},
\end{equation}

where $N"$ is the new signal size after zero-padding in Fourier space. We have thus zoomed the image by a factor $z=(N"/N')$. Note that if we only zero-pad the image in Fourier space, as it is often done, then $N'=N$, which severely constrains the zooming factor to $z=N"/N$.

Considering a coronagraphic image $\lambda_1$ that we want to resample at a smaller wavelength $\lambda_0$ , we need to satisfy the following equation:

\begin{equation}
  \label{eq-choice-N-prime-N-second}
  \frac{N'}{N"} = \frac{\lambda_1}{\lambda_0} .
\end{equation}

Having two degrees of freedom $N'$ and $N"$ allows us, in practice, to satisfy this equality with a good precision (typically $10^{-4}$) for reasonable values of $N'$ and $N"$ for all wavelengths of the IRDIS or IFS instruments.

\bsp	
\label{lastpage}
\end{document}